\documentclass[12pt]{iopart}

\def\sqrtsNN{\mbox{$\sqrt{s_\mathrm{NN}}$}}
\usepackage{graphicx}
\begin{document}

\title[Incident-energy and system-size dependence of directed flow]{Incident-energy and system-size dependence of directed flow}

\author{Gang Wang (for the STAR\footnote{For the full list of STAR authors and acknowledgements, see appendix `Collaborations' in this volume.} Collaboration)}

\address{University of California,
Los Angeles, California 90095, USA}
\ead{gwang@physics.ucla.edu}

\begin{abstract}
We present STAR's measurements of directed flow for charged
hadrons in Au+Au and Cu+Cu collisions at $\sqrtsNN = 200$ GeV and $62.4$ GeV, 
as a function of pseudorapidity, transverse momentum and
centrality. We find that directed flow depends on the incident energy, but not on the system size. 
We extend the validity of limiting fragmentation hypothesis to different collision systems.
\end{abstract}


\section{Motivation}
Directed flow is quantified by the first harmonic ($v_1$) in the Fourier expansion of the azimuthal distribution of
produced particles with respect to the reaction plane~\cite{Methods}. It describes the collective sideward motion 
of produced particles and nuclear fragments, and carries information from the very early stage of the collision~\cite{Whitepaper}. 
Charged-hadron $v_1$ results at RHIC have been previously reported over a wide range of incident energies in  
Au+Au collisions~\cite{v1v4,PHOBOSv1,v1ZDCSMD1,v1ZDCSMD2}, and they support the limiting fragmentation hypothesis~\cite{PHOBOSv1,v1ZDCSMD1,LimFr}.
In RHIC run V (2005), lighter nuclei were collided at $\sqrtsNN = 200$ GeV and $62.4$ GeV, which enables us to study the
system-size dependence of directed flow, as well as the incident-energy dependence.

\section{Data sets and the study}
This study is based on eight million 200 GeV Au+Au events, five million 62.4 GeV Au+Au events, twelve million 200 GeV Cu+Cu
events, and eight million 62.4 GeV Cu+Cu events. All of them were obtained from a minimum-bias trigger. All errors
are statistical. Tracks of charged particles are reconstructed by STAR's main TPC~\cite{TPC} and forward TPCs~\cite{FTPC},
with the pseudorapidity ($\eta$) coverage of $|\eta|<1.3$ and $2.5 < |\eta| < 4.0$, respectively.
The centrality definition and cuts used are the same as in Ref.~\cite{Centrality}. 

\begin{table}[hbt]
\caption{The resolution of the 1st-order full event plane provided by STAR
ZDC-SMDs, as determined from the sub-event correlation between east and west SMDs. 
The errors in the table are statistical.}
\footnotesize\rm
\begin{center}
\begin{tabular}{c|c|c|c|c}\hline \hline
Centrality &  200 GeV Au+Au & 62.4 GeV Au+Au & 200 GeV Cu+Cu & 62.4 GeV Cu+Cu
\\ \hline
$70\%-80\%$  & $0.296\pm 0.003$& $0.179\pm 0.005$ &
\\ \hline
     $60\%-70\%$ & $0.348\pm 0.003$ & $0.185\pm 0.004$ & 
\\ \hline
     $50\%-60\%$  & $0.382\pm 0.002$& $0.176\pm 0.005$& $0.140\pm 0.003$ & $0.039\pm 0.011$
\\ \hline
 $40\%-50\%$      & $0.397\pm 0.002$ & $0.167\pm 0.005$& $0.144\pm 0.003$ & $0.043\pm 0.011$
\\ \hline
 $30\%-40\%$    & $0.390\pm 0.002$ & $0.138\pm 0.006$& $0.147\pm 0.003$ & $0.032\pm 0.015$
\\ \hline
  $20\%-30\%$   & $0.365\pm 0.002$ & $0.110\pm 0.008$& $0.121\pm 0.004$
\\ \hline
 $10\%-20\%$    & $0.309\pm 0.003$ & $0.081\pm 0.010$& $0.095\pm 0.005$
\\ \hline
 $5\%-10\%$     & $0.220\pm 0.006$&                  & $0.059\pm 0.008$  
\\ \hline
 $0-5\%$        & $0.120\pm 0.002$&			& $0.059\pm 0.008$
\\    \hline \hline
\end{tabular}
\end{center}
\label{tbl:resolution}
\end{table}

At RHIC energies, it is a challenge to measure $v_1$ accurately due to the small signal of $v_1$ itself and 
the large systematic error arising from non-flow correlations.
In this talk, STAR's preliminary $v_1$\{ZDC-SMD\} results are obtained with non-flow effects minimized
by determining the event plane from spectator neutrons, the same approach as used in Ref.~\cite{v1ZDCSMD1,v1ZDCSMD2}.
The resolution, as defined in Ref.~\cite{Methods}, of the first-order event plane reconstructed
with STAR ZDC-SMDs~\cite{ZDC-SMD} is listed in Table~\ref{tbl:resolution}.
The magnitude of the event plane resolution increases with the spectator $v_1$ and the detector efficiency of the ZDC-SMDs, 
and the latter is smaller for a lower incident energy and/or a smaller collision system.
The method $v_1$\{ZDC-SMD\} fails in some centralities, where the event plane resolution is consistent with zero.

\section{Results}
Figures in this talk follow the same convention as in Ref.~\cite{v1ZDCSMD1}.

\begin{figure}
\begin{minipage}[t]{77mm}
\includegraphics[width=1.0\textwidth]{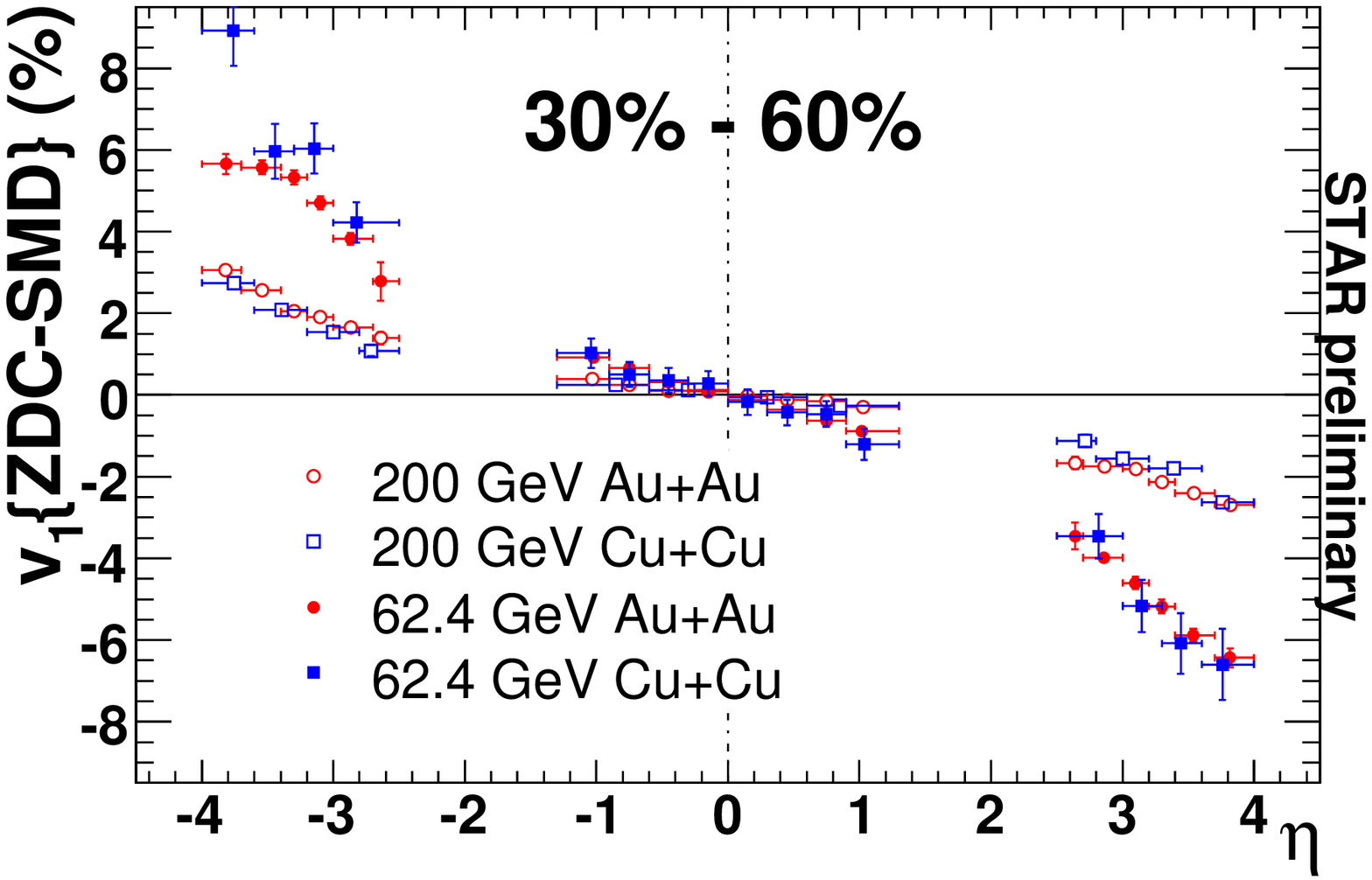}
\hspace{-2cm}
\caption{Charged-hadron $v_1$ vs. $\eta$, in Au+Au and Cu+Cu collisions at 200 GeV and 62.4 GeV.
The plotted errors are statistical.}
\label{fig:v1_eta}
\end{minipage}
\hfill
\begin{minipage}[t]{77mm}
\includegraphics[width=1.0\textwidth]{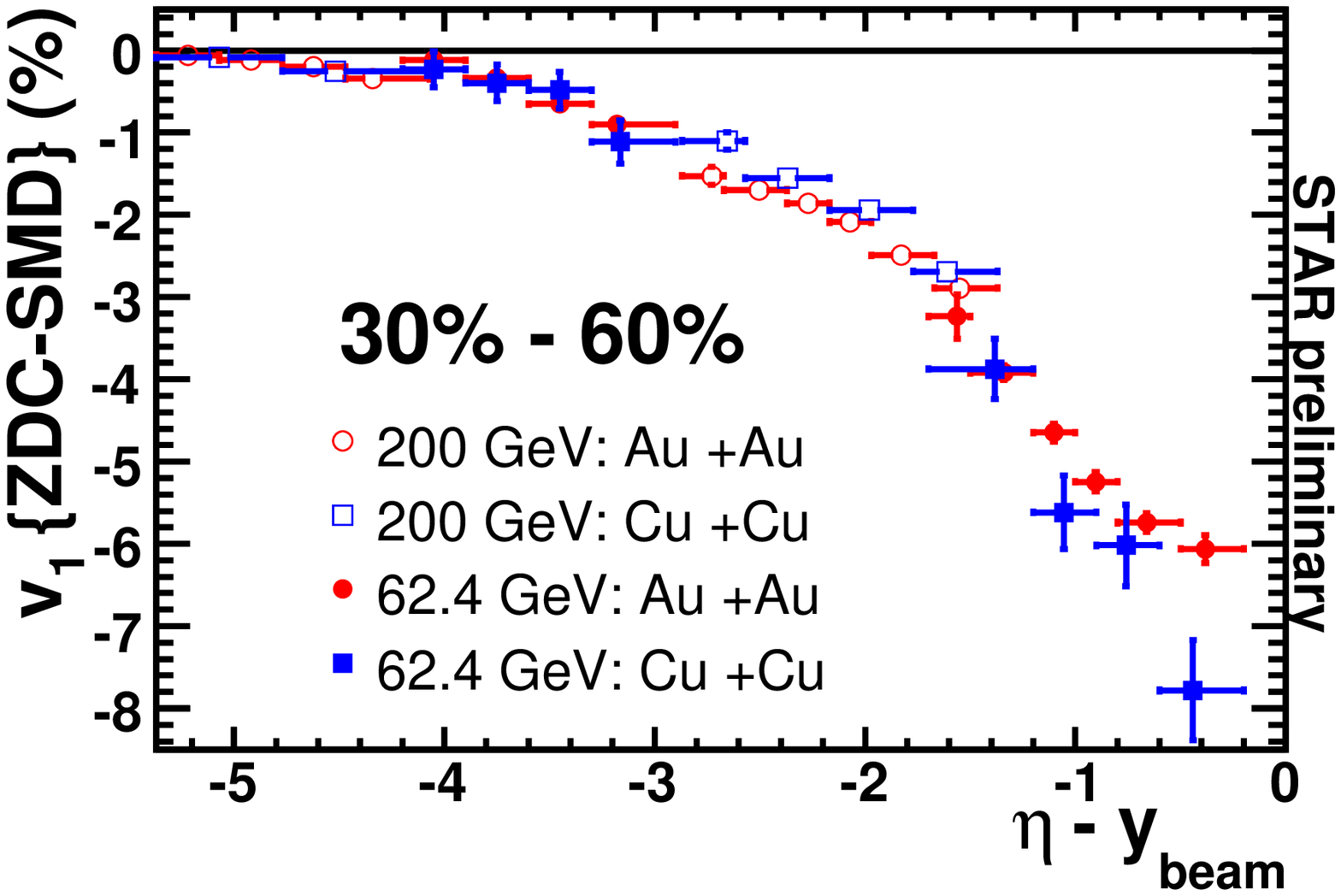}
\caption{Charged-hadron $v_1$ vs. $\eta - y_{\rm beam}$, in Au+Au and Cu+Cu collisions at 200 GeV and 62.4 GeV.
The plotted errors are statistical.}
\label{fig:v1_LimFr}
\end{minipage}
\end{figure}

Fig.~\ref{fig:v1_eta} shows $v_1$ as a function of pseudorapidity, $\eta$, in $30\% - 60\%$ most central Au+Au and 
Cu+Cu collisions at $\sqrtsNN = 200$ GeV and $64.2$ GeV. 
Data points in Fig.~\ref{fig:v1_eta} fall into two bands, and the higher band corresponds to the lower energy, which
reveals the incident-energy dependence of $v_1(\eta)$.
For the pseudorapidity and centrality range we studied, $v_1(\eta)$ in Au+Au collisions and Cu+Cu collisions are consistent 
within errors if both of them were obtained at the same incident energy. 
It is interesting that $v_1$ does not change from Au+Au collisisons to Cu+Cu collisions,
while the system size is reduced by $1/3$. 
A good consistency of $v_1$ between Au+Au and Cu+Cu collisions is observed even
for the region near midrapidity, where $v_2$ in Cu+Cu collisions is considerably lower than 
that in Au+Au collisions~\cite{v2_scaling}.
Unlike $v_2/\epsilon$ which scales with $\frac{1}{S} \frac{dN_{ch}}{dy}$ (interpreted to be
the participant density~\cite{v2_scaling} or the system length~\cite{Aihong}), 
$v_1$ is found to be independent of the system size. Instead, it scales with the
incident energy. A possible explanation to the different scalings for
$v_2/\epsilon$ and $v_1$  might come from the way in which they are
developed : to produce $v_2$, intensive momentum exchanges among particles are
needed (and the number of momentum exchanges is related to the participant density or the system length),
while to produce $v_1$, one in principle needs only different rapidity losses
(related to the incident energy) for particles with different
distances to the center of the participant zone.

In Fig.~\ref{fig:v1_LimFr}, we re-arrange the data points of Fig.~\ref{fig:v1_eta} in the
projectile frame, where zero on the horizontal axis corresponds to the beam rapidity, $y_{\rm beam}$,
for each of the incident energies. Within three units from $y_{\rm beam}$, most data points
fall into a universal curve of $v_1$ as a function of $\eta - y_{\rm beam}$.
This incident-energy scaling behavior of directed flow
suggests that the limiting fragmentation hypothesis~\cite{LimFr}
holds even for different collision systems.

\begin{figure}
\begin{minipage}[t]{77mm}
\includegraphics[width=1.0\textwidth]{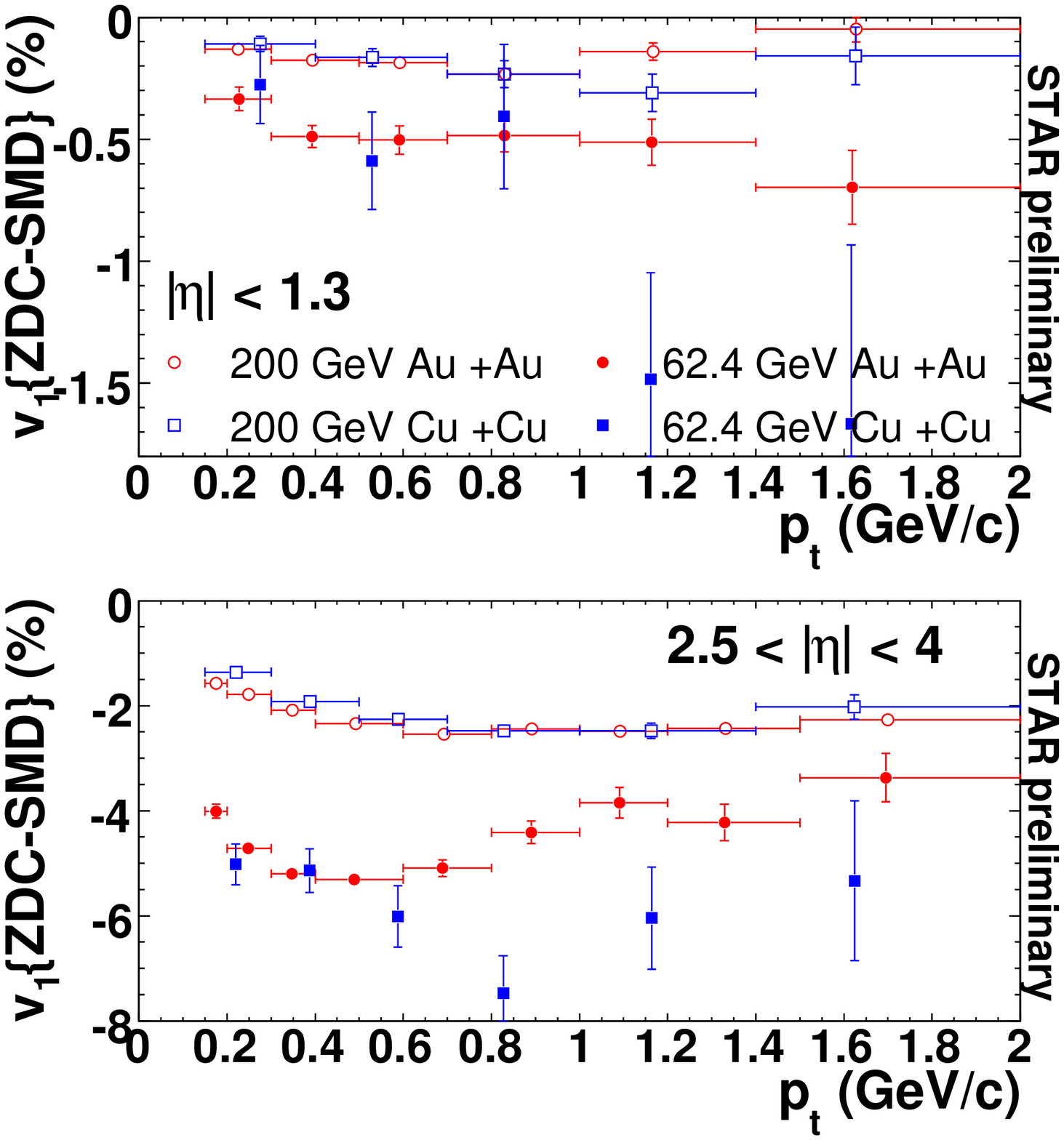}
\hspace{-2cm}
\caption{Charged-hadron $v_1$ vs. $p_t$, in $30\% - 60\%$ Au+Au and Cu+Cu collisions at 200 GeV and 62.4 GeV. The upper panel shows the results measured in the main TPC, and the lower panel, in the forward TPCs.
The plotted errors are statistical.}
\label{fig:v1_pt}
\end{minipage}
\hfill
\begin{minipage}[t]{77mm}
\includegraphics[width=1.0\textwidth]{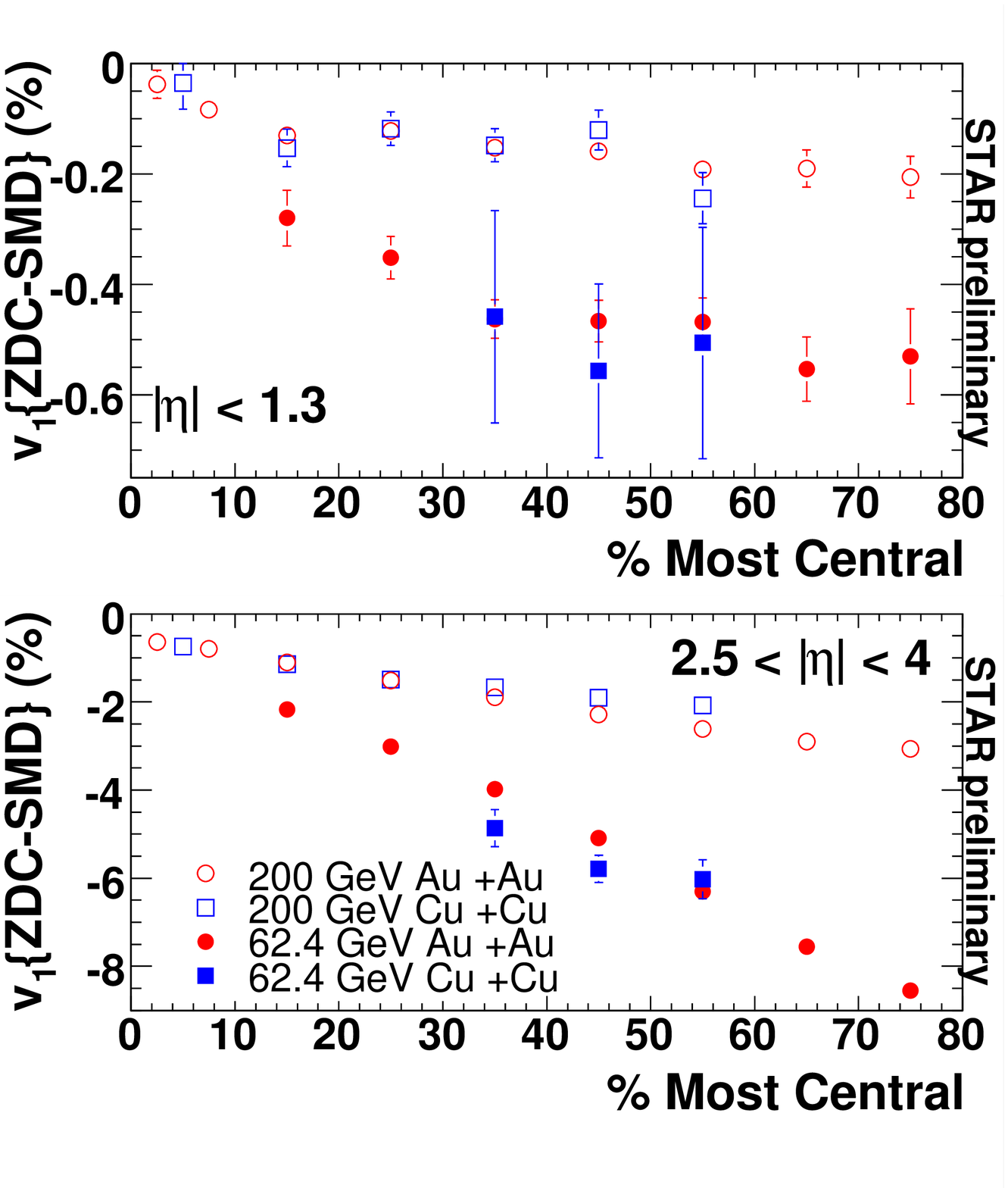}
\caption{Charged-hadron $p_t$-integrated $v_1$ vs. centrality, in Au+Au and Cu+Cu collisions at 200 GeV and 62.4 GeV. The upper panel shows the results measured in the main TPC, and the lower panel, in the forward TPCs.
The plotted errors are statistical.}
\label{fig:v1_int}
\end{minipage}
\end{figure}

The transverse-momentum ($p_t$) dependence of $v_1$ for charged hadrons is shown in Fig.~\ref{fig:v1_pt}.
The upper panel shows the results measured in the main TPC, and the lower panel, in the forward TPCs.
Since $v_1(\eta, p_t)$ is asymmetric about $\eta=0$,
the integral of $v_1(\eta, p_t)$ over a symmetric $\eta$ range goes to zero.
We change $v_1(\eta, p_t)$ of particles with negative $\eta$
into $-v_1(-\eta, p_t)$, and integrate over all $\eta$.
At 200 GeV, $v_1(p_t)$ in Au+Au and Cu+Cu collisions are very close to each other 
(better visible in the forward pseudorapidity region),
and $v_1(p_t)$ saturates above $p_t \approx 0.8$\,GeV/$c$.
At 62.4 GeV, $v_1(p_t)$ reaches its maximum in magnitude at
$p_t \approx 0.6$ GeV/$c$, then saturates ($| \eta | < 1.3$) or slightly decreases ($2.5 < | \eta | < 4.0$).
Within statistical errors, the incident-energy scaling holds
in $v_1(p_t)$ as well.

Fig.~\ref{fig:v1_int} shows charged-hadron $p_t$-integrated $v_1$ as a function of centrality 
for the two collision systems at 200 GeV and 62.4 GeV,
and the results are presented in two panels defined in the same way as  Fig.~\ref{fig:v1_pt}.
On the whole, the magnitude of integrated $v_1$ approaches zero in central collisions and increases
as the collisions become more peripheral.
At 200 GeV, the $p_t$-integrated $v_1$ in Au+Au collisions is consistent with that in Cu+Cu collisions 
in almost all centrality bins,
in regions both near the midrapidity ($| \eta | < 1.3$) and away from it ($2.5 < | \eta | < 4.0$).
This system-size independence is seemingly true at 62.4 GeV.
We can see that the incident-energy scaling of $v_1$ works well as a funtion
of centrality, instead of the participant density, with which the $v_2/\epsilon$ scales~\cite{v2_scaling}.
This indicates that the initial geometrical shape, but not the density of produced particles, plays an
important role in the formation of directed flow.

\end{document}